# High-frequency nanotube mechanical resonators

J. Chaste, M. Sledzinska, M. Zdrojek, J. Moser, and A. Bachtold

CIN2(ICN-CSIC), Catalan Institute of Nanotechnology, Campus de la UAB, 08193 Bellaterra

(Barcelona), Spain

We report on a simple method to fabricate high-frequency nanotube mechanical resonators reproducibly. We measure resonance frequencies as high as 4.2 GHz for the fundamental eigenmode and 11 GHz for higher order eigenmodes. The high-frequency resonances are achieved using short suspended nanotubes and by introducing tensile stress in the nanotube. These devices allow us to determine the coefficient of the thermal expansion of an individual nanotube, which is negative and is about $-0.7 \cdot 10^{-5}$ 1/K at room temperature. High-frequency resonators made of nanotubes hold promise for mass sensing and experiments in the quantum limit.

Mechanical resonators have attracted considerable interest as ultra-sensitive detectors of mass [1-5] and force [6-8], and as macroscopic objects can be cooled to the quantum motion limit [9-11]. A key parameter in these experiments is the resonance frequency $f_0$, which is often desirable to have as high as possible. For instance, a high resonance frequency is expected to improve the sensitivity of mass sensing; it also makes the temperature at which quantum phenomena appear within reach of standard refrigerators. Flexural motion at $f_0$ higher than 1 GHz was detected using single- and doubly-clamped microfabricated resonators [12,13]. Single-wall carbon nanotubes have allowed the fabrication of nanoelectromechanical resonators endowed with excellent properties [2-5,8,14-20], but the highest reported resonance frequency is rather modest (below 600 MHz).

In this Letter, we report on a simple and reliable method to fabricate high-frequency nanotube resonators where $f_0$ can be as high as 4.2 GHz for the fundamental eigenmode and 11 GHz for higher order eigenmodes. The high-frequency resonators are achieved using a device layout in which the suspended nanotube is short. The resonance frequency is further increased by lowering the temperature $T$. We attribute the latter behavior to the thermal expansion of the metal electrodes (used to clamp the nanotube), which increases the tensile stress in the nanotube upon lowering $T$.

Nanotube resonators were fabricated using conventional nanofabrication techniques. Nanotubes were grown by chemical vapor deposition on a highly-resistive silicon wafer coated with a 1 μm thick oxide layer. In order to achieve high-frequency resonators, care was taken to (1) design contact electrodes with a short separation ranging from 100 to 700 nm and (2) select straight segments of nanotubes by atomic force microscopy (AFM) to minimize slack once the nanotube is suspended. We used electron-beam lithography and Cr/Au evaporation to pattern contact electrodes as well as a side-gate electrode (Fig. 1a). (For those devices whose electrode separation is more than 600 nm, we used a highly-doped Si wafer as a backgate.) The nanotubes were suspended by etching ~250 nm of the oxide with fluoridic acid.

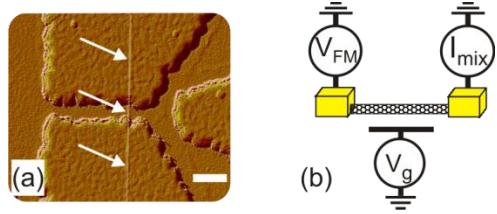

Fig. 1 (a) AFM image of the device prior to removal of the supporting oxide. The scale bar is 300 nm. The white arrows point to the nanotube. (b) Schematic of the actuation/detection setup. A frequency modulated voltage $V_{FM}$ is applied to the device. The motion is detected by measuring the mixing current $I_{mix}$.

Figures 2a,b show two mechanical resonances at 4.2 and 11 GHz for a device whose contact electrodes are ~110 nm apart. The motion was driven and detected using the frequency modulation (FM) mixing technique [18] (Fig. 1b). The resonances consist of a central peak flanked by two lobes, which is consistent with the typical lineshape of a mechanical resonance in a FM measurement [18] (the observed asymmetry in the resonances is probably due to the Duffing force). The properties of the resonator slightly changed during the X cool-downs experienced by the device. That is, the lower $f_0$ fluctuates between 3.8 and 4.3 GHz and the higher $f_0$ between 11 and 11.2 GHz.

These resonances correspond to flexural eigenmodes, since other eigenmodes are expected to have much larger resonance frequencies ($f_0 \approx$ 240 GHz for the lowest longitudinal and twisting eigenmodes of a 110 nm long nanotube [21]). We assign the 4.2 GHz resonance to the fundamental eigenmode, since it is the lowest in frequency and shows the largest current signal at resonance. Accordingly, the 11 GHz resonance corresponds to a higher-order eigenmode, which is likely the 3rd mode, since the 2nd mode is expected to be driven weakly due to the symmetry of the device (the gate electrode is symmetrically placed with respect to the nanotube). The ratio between the 11 and the 4.2 GHz frequencies is rather close to 3, the value expected for a nanotube under tension. One reason why the ratio is not exactly 3 could be related to the influence of the clamping electrodes (I. Wilson-Rae, private communication).

We confirmed the mechanical origin of these resonances by depositing xenon atoms onto the nanotube in situ and observing a drop in the resonance frequencies (Fig. 2e). The reason for the drop in $f_0$ is that Xe atoms adsorb on the nanotube surface and thus increase the mass of the mechanical resonator (these results will be further discussed elsewhere). Another signature of the mechanical origin is the strong temperature dependence of $f_0$ (Fig. 1e); we indeed studied dozens of mechanical resonators with the FM mixing technique and occasionally observed some low-amplitude resonances whose origin was purely electrical, but whose $f_0$ would not change with temperature.

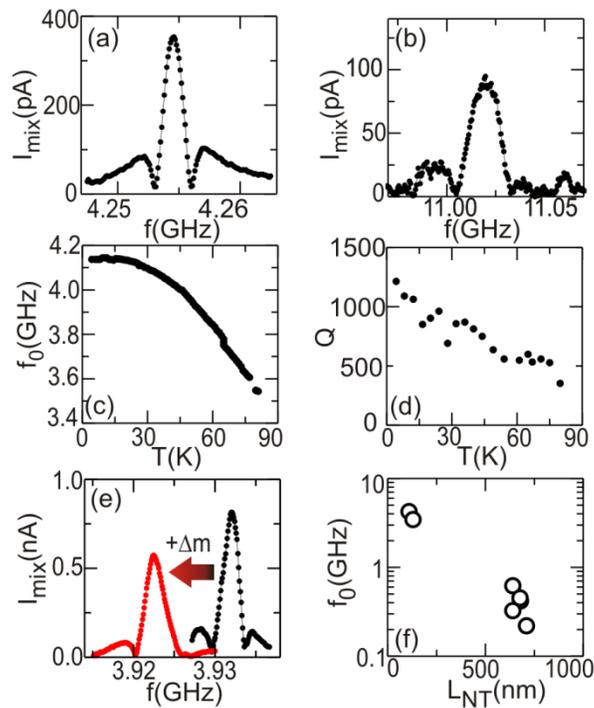

Fig. 2 (a,b) Mechanical resonances measured at 4 K. The $V_{FM}$ amplitude is 5 mV in (a) and 10 mV in (b). (c,d) Temperature dependence of the resonance frequency and the quality factor for the resonance in (a) measured with $V_{FM}$ = 4 mV. (e) Frequency shift of the resonance after having added Xe atoms onto the resonator. Xe atoms were admitted in the sample chamber from a gas reservoir at 300 K using a pinhole doser. (f) Resonance frequency as a function of the nanotube length at 4 K for different resonators. [JULIEN: Double check les valeurs de $V_{FM}$]

Figure 2d shows that the quality factor $Q$ increases upon lowering temperature and is as high as 1200 at 4 K and for a driving excitation $V_{FM}$ = 4 mV. We emphasize that the quality factor varies with the driving excitation ($Q$ increases by a factor ~3 upon increasing $V_{FM}$ from 1 to 10 mV). It is not clear whether this variation is due to nonlinear damping [8,20] or is related to Joule heating (which is induced by the large applied voltage).

Two important parameters that allow us to tune the resonance to high frequency are the length of the suspended nanotube segment ($L_{NT}$) and the temperature. Indeed, Fig. 2f displays the resonance frequency of the fundamental eigenmode at 4 K for the 7 resonators that were fabricated with the method described above; the resonance frequency increases as $L_{NT}$ decreases, as expected for mechanical resonators. In addition, the resonance frequency dramatically increases upon lowering $T$ as shown in Fig. 2c and in the upper curve in Fig. 3a. The latter behavior is related to the $T$ dependence of the tensile stress within the nanotube: in our device layout (inset of Fig. 3a) the suspended parts of the clamping Au electrodes contract upon lowering $T$, thus they increase the tensile stress within the nanotube (the estimated contraction [22] of Au from 300 to 4K is $|\Delta l_{Au}| \approx$ 1.6 nm and is larger than that of Si [23] which is $|\Delta l_{Si}| \approx$ 0.2 nm in Fig. 3a). By contrast, the resonance frequency is weakly sensitive to the voltage applied on the gate (see supplementary information).

From the $T$ dependence of $f_0$, we evaluate the elongation $\Delta l$ imposed on the nanotube using the relation

$$f_0 = 0.5\sqrt{T_0 / mL_{NT}} \qquad\qquad (1)$$

valids for a beam under tensile stress and using Hooke´s law $\Delta l = T_0 L_{NT} / E\pi(r^2 - (r - 0.167nm)^2)$ for a hollow cylinder with built-in tension $T_0$. Here $m$= 2.6 ag is the mass of the nanotube, $L_{NT}$=640 nm its length, $r$=0.85 nm its radius, and $E$=1.25 TPa its Young's modulus ($L_{NT}$ and $r$ were measured with AFM before the nanotube suspension). Figure 3b shows that the elongation and its variation with $T$ are very small (of the order of 100 pm).

We are now in a position to estimate the thermal expansion coefficient (TEC) of an individual nanotube, a property that has not been measured thus far. For this, we use the definition of the TEC

$$\alpha_{NT} = \frac{\Delta l_{NT}}{\Delta T}\frac{1}{L_{NT}} \qquad\qquad (2)$$

together with the hypothesis of the conservation of length $\Delta l_{NT} = \Delta l_{Si} - \Delta l_{Au} - \Delta l$. Figure 3c shows that the TEC of the nanotube is negative and that its magnitude is rather large ($10^{-6} - 10^{-5}$ 1/K). We obtain similar results for the other nanotubes (for which $f_0$ could be measured from 300 to 4 K). Interestingly, this TEC is similar to what was recently measured in graphene [24,25] and also to the TEC estimated using numerical simulations on unclamped nanotubes [26].

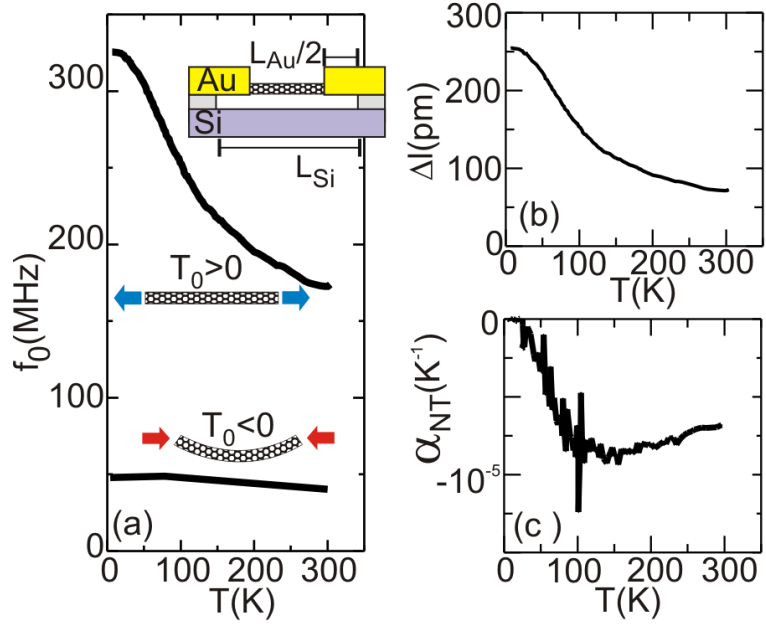

Fig. 3(a) Resonance frequency as a function of $T$ for a 640 nm long nanotube (upper curve, the built-in tension $T_0$ is positive). For purposes of comparison, the lower curve corresponds to the measurement of the lowest resonance of a nanotube with slack ($T_0 < 0$) whose length is about 1 μm (the device is fabricated as in Ref. [15]).The upper inset shows the schematic of the nanotube resonator. (b) Elongation imposed on the nanotube under tensile stress, obtained using $L_{Au}$=520 nm and $L_{Si}$=1160 nm ($L_{Au}$ is taken as half of the etched oxide depth measured with AFM). (c) Thermal expansion coefficient of the nanotube in (b).

In conclusion, we have reported on a simple method to fabricate high-frequency nanotube resonators under tensile stress. The tensile stress increases upon lowering temperature, since the suspended Au electrodes contract. High-frequency resonators made from nanotubes hold promise for various scientific and technological applications, such as mass sensing and experiments in the quantum limit. Finally, our device layout allows us to estimate the TEC of an individual nanotube. In future high-speed, high-density nanotube circuits, a knowledge of the nanotube TEC will help address thermal management issues.

We acknowledge support from the European Union (RODIN, FP7), the Spanish ministry (FIS2009-11284), and the Catalan government (AGAUR, SGR).

# High-frequency nanotube mechanical resonators

J. Chaste, M. Sledzinska, M. Zdrojek, J. Moser, and A. Bachtold

CIN2(ICN-CSIC), Catalan Institute of Nanotechnology, Campus de la UAB, 08193 Bellaterra

(Barcelona), Spain

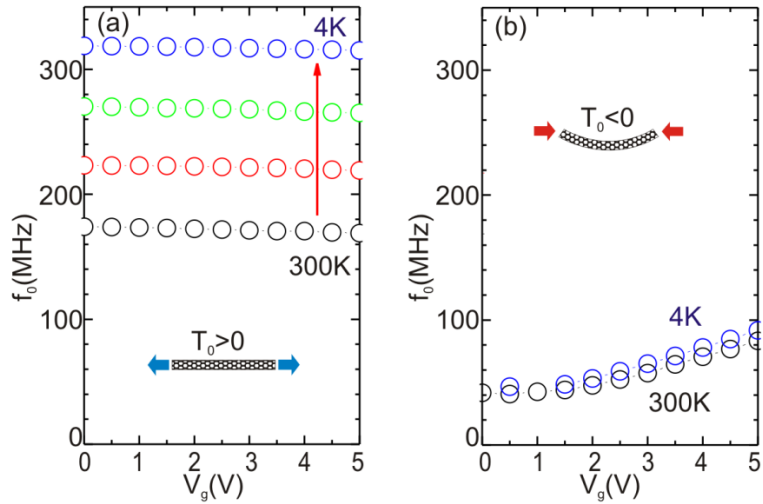

Fig. 1S(a) Resonance frequency of a nanotube under tensile stress as a function of $V_g$ for various $T$ (4, 91, 153, and 300K). (b) Same measurement as in (a) for a nanotube with slack. We present only the data of the lowest measured resonance. The nanotubes in (a) and (b) are the same as those in Fig. 3a.

Figure 1Sa shows the resonance frequency as a function of $V_g$ (applied on the backgate) at different temperatures and for a device fabricated with the process described in the main text and having contact electrodes separated by ~640 nm. The resonance frequency is weakly sensitive to $V_g$. This $V_g$ dependence of $f_0$ differs greatly from what is measured in nanotube resonators with slack (Fig. 1Sb) where $\delta f_0 / f_0$ is much larger and positive. The latter behavior is well-documented and is attributed to the tension $T_e$ that builds in the nanotube as it bends towards the backgate upon increasing $V_g$ [1,2,3]. The fact that this behavior is not observed in the resonators fabricated with the process described in this paper (Fig. 1Sa) is an indication that the nanotube is under tensile stress and that the built-in tension $T_0$ is much larger than $T_e$ ($T_0 \gg T_e$ since $f_0 \propto \sqrt{T_0 + T_e}$ for a beam under tensile stress and that $f_0$ is not affected by the tension induced by the electrostatic force).